A causal machine learning approach to disentangling the pressure-satisfaction pathways in depression

Zhaojin Nan[1]

1 Department of Statistics and Data Science, Washington University in St. Louis, St. Louis, MO. 63130, USA

## Abstract

Purpose: Prior research has established perceived pressure and life satisfaction as important correlates of depression, yet their causal interplay remains insufficiently identified. This study aims to disentangle whether satisfaction acts as an independent protective factor or operates by buffering pressure, and to identify population-specific risk profiles across students and workers.

Methods: We applied a causal machine learning framework to harmonized data from India, China, and Malaysia (total N=28,243). We integrated random forests and logistic regression with Causal Mediation Analysis and Causal Forests. To resolve theoretical ambiguity regarding directionality, we validated the causal pathway between pressure and satisfaction using numerical simulation benchmarks.

Results: Pressure emerged as the dominant predictor across all cohorts. Simulation-validated analysis confirmed a causal pathway flowing from Life Satisfaction → Pressure → Depression, rejecting the reverse hypothesis. Satisfaction mitigated depression partially through pressure reduction (proportion mediated ≈ 15.1%), rather than functioning exclusively as a direct mechanism. A distinct developmental reversal was observed: younger age predicted vulnerability in students, whereas older age predicted risk in workers. Causal forests further revealed that the depressogenic impact of pressure was significantly amplified in students with high anxiety.

Conclusion: Pressure acts as the proximal bottleneck for depression risk, while life satisfaction functions as an antecedent buffer. These findings challenge uniform risk models by highlighting age-related context dependence and suggest that precision interventions targeting stress reduction and high-anxiety subgroups offer the most effective pathway for breaking the causal cycle of depression.

## Highlights:

1 Simulation validates that satisfaction buffers depression by reducing pressure.

2 Perceived pressure emerges as the dominant depression predictor across all cohorts.

3 Risk profiles reverse with younger students and older workers being most vulnerable.

4 Causal forests reveal that high anxiety significantly amplifies pressure’s impact.

# 1 Introduction

Depression has emerged as a predominant public health challenge globally, with incident cases rising by approximately 59.3 % between 1990 and 2019 (Wu et al., 2024). This burden is particularly acute in Asia. In 2019 alone, India and China reported over 54 million and 41 million incident cases of depressive disorders, respectively (Wu et al., 2024). While the proliferation of machine learning (ML) has significantly advanced our ability to predict depression risk with high accuracy, a critical gap remains between prediction and intervention. Standard predictive models often identify key correlates, such as age, frailty, social support, marital status, physical health, sleep quality, and socioeconomic conditions (Duan et al., 2020; Liu et al., 2021； Islam et al., 2022; Li et al., 2024). Studies from China, India, and Bangladesh further highlight the roles of educational context, employment status, family environment, and financial stressors in shaping depression risk (Lowe et al., 2009; Singh et al., 2017). But these models often fail to elucidate the underlying causal mechanisms required to inform targeted treatment strategies (Abécassis et al., 2025).

A central unresolved question in the etiology of depression concerns the mechanistic relationship between perceived pressure and life satisfaction. Both are well-established determinants of mental health. Longitudinal studies indicate that life dissatisfaction predicts the onset of major depressive disorder (Rissanen et al., 2011; Almeida et al., 2021; Ooi et al., 2022; Arslan et al., 2025), while separate lines of research identify chronic time pressure and perceived stress as potent drivers of depressive symptoms (Wang et al., 2024; Ogden et al., 2025). However, most empirical studies rely on cross-sectional regression or moderation frameworks that do not distinguish mediation from direct effects under explicit causal assumptions. Existing literature often treats them as independent correlates or relies on assumed, bidirectional associations without rigorous testing (Lee et al., 2016). Consequently, it remains unclear whether life satisfaction acts as an independent protective factor, or if it functions primarily as an antecedent buffer that mitigates depression by lowering perceived pressure. Resolving this directionality is clinically vital: if pressure is the proximal driver, interventions must prioritize stress reduction over broad well-being promotion to be effective.

Furthermore, the impact of demographic risk factors, particularly age, appears inconsistent across the literature. While some studies suggest depression prevalence peaks in adolescence and young adulthood (Brody et al., 2025), others identify elevated risks in older populations associated with frailty and declining physical health (Lin et al., 2022). These inconsistencies may stem from contextual confounding, as student and working populations are often examined in isolation rather than within a unified analytic framework. Most prior research (e.g. Bonde, 2008; Besharat et al., 2014; Worrall et al., 2020; Islam et al., 2022) focuses on specific isolated populations, either students or workers, making it difficult to systematically observe how these risk profiles diverge across life stages. Disentangling such population-specific effects is critical for understanding how structural and psychosocial pressures shape depression across the life course.

To address these gaps, this study employs a causal machine learning approach to disentangle the pathways linking pressure, satisfaction, and depression across diverse Asian populations. Unlike traditional regression analyses, we integrate Causal Forests (Athey et al., 2019) and Causal Mediation Analysis (CMA) (Imai et al., 2010) with predictive modeling (Random Forests). We aim to (1) rigorously determine whether pressure acts independently or mediates the effect of life satisfaction on depression; (2) use causal forests to test if these effects remain stable across individuals or vary by subgroup; and (3) systematically analyze harmonized data from students (India, Malaysia) and workers (China, India) to identify population-specific risk profiles, specifically examining the divergent role of age. By moving beyond risk prediction to causal explanation, this study offers statistical evidence to guide more precise, stress-focused interventions in academic and occupational settings.

# 2 Materials and methods

## 2.1 Study populations and data sources

This study utilized three distinct datasets from India (Sharma et al., 2025), Malaysia (Islam, 2023), and China (Zhao et al. 2014) to systematically examine the mechanisms of depression across diverse cultural and developmental contexts. China and India were selected as the primary analytical foci because they bear the highest absolute burden of depressive disorders globally in the general population as well as among adolescents. The inclusion of Malaysia alongside India and China ensures representation across the three major sub-regions of Asia, spanning South Asia (India), East Asia (China), and Southeast Asia (Malaysia), allowing for a more robust assessment of whether risk mechanisms remain stable across distinct cultural and socioeconomic environments. These datasets were specifically chosen to enable a rigorous comparison between students and workers. The India dataset serves as a heterogeneous "bridge" cohort containing both populations, while the Malaysia (student-only) and China (worker-only) datasets serve as specialized external validation cohorts. This design allows us to test the consistency of our causal findings both within a single country and across national borders.

Data for the Indian cohort were derived from an anonymous cross-sectional survey conducted between January and June 2023 (Sharma et al., 2025). This dataset comprises 2,556 individuals from diverse professional and social backgrounds across multiple cities. Uniquely, this dataset includes both working professionals and students, allowing for internal comparative analysis within a single cultural setting. Depression was assessed as a binary outcome, with 455 participants (approximately 18%) reporting positive symptoms. The survey includes a set of 24 variables, including demographics (age, gender, city), academic and occupational stressors (e.g., academic pressure, job satisfaction, sleep duration), lifestyle factors (diet, work/study hours), and mental health indicators (e.g., suicidal ideation, family history of mental illness).

To provide an external validation benchmark for the student population, we utilized a focused dataset from the International Islamic University Malaysia. Collected via a Google Form survey, this Malaysian dataset includes responses from 101 university students and 12 variables (Islam, 2023). It serves as a specific comparator for academic stress factors, capturing variables such as. demographic details (age, gender), academic information (course of study, current grade point average (CGPA), academic year), and mental health indicators (e.g., anxiety, panic attacks, and prior mental health treatment). The prevalence of depressive symptoms in this cohort was 35%.

Data for the working population in China were drawn from the Harmonized China Health and Retirement Longitudinal Study (CHARLS), a nationally representative survey funded by the National Institute on Aging (Zhao et al. 2014). We utilized the most recent available wave, comprising 25,586 respondents and 3,400 variables. This extensive dataset focuses on older adults and the working-age population, providing granular detail on demographic information (age, gender), physical and mental health status (e.g., self-reported memory, life satisfaction, sleep duration), lifestyle habits (physical activity, work hours), and financial conditions (income, loans, housing wealth, and financial stress). Approximately 20% (n=5,090) of the respondents reported symptoms of depression. Table 1 summarizes the key characteristics of these three datasets.

Table 1 Summary of study cohorts. GPA: current grade point average; CHARLS: the Harmonized China Health and Retirement Longitudinal Study

| Dataset | Population type | Sample Size | Depression Prevalence | Number of Features | Key Features | Source |
|---|---|---|---|---|---|---|
| India | Students & Workers | 2556 | 18% (n=455) | 24 | Demographics, stressors, lifestyle, mental health history | Kaggle (Sharma et al., 2025) |
| Malaysia | Students | 101 | 35% (n=35) | 12 | Demographics, academic stress, CGPA, anxiety | Kaggle (Islam, 2023) |
| China | Workers | 25586 | 20% (n=5090) | 3400 | Demographics, health status, lifestyle, finances | CHARLS (Zhao et al. 2014) |

## 2.2 Variable harmonization and preprocessing

To ensure consistency and comparability across the three national datasets, a rigorous harmonization procedure was implemented to align diverse measurement scales into a unified feature set. A common set of core variables was first identified across all cohorts, including gender, age, academic performance (for students), perceived pressure, workload, financial stress, mental health history, sleep quality, life satisfaction, social support, future concern, anxiety (for students), and extracurricular (for students) or physical activity (for workers). Because the original datasets served different primary populations, with Malaysia consisting exclusively of students and China comprising only working adults, the heterogeneous Indian dataset was explicitly stratified into two distinct subgroups: "India students" and "India workers," based on the respondents' reported professional status. Subsequent cleaning and imputation were performed separately on student-oriented datasets (India students and Malaysia) and working-adult datasets (India workers and China).

Following stratification, all variables were standardized to a uniform encoding scheme. The dependent variable, depression, was binary encoded across all datasets, with positive symptoms recoded as 1 and absence of symptoms as 0. Demographic and historical predictors, such as gender and family history of mental illness, were similarly binary encoded. To harmonize predictors measured on disparate scales, such as academic performance (originally reported as CGPA) and financial stress (originally reported via various loan indicators), we transformed these continuous or multi-level variables into standardized ordinal scales ranging from 0 to 5. Perceived pressure, workload, sleep quality, life satisfaction, social support, extracurricular/physical activity, and future concern were likewise normalized to this 0-5 scale to prevent scalar bias during model training. Anxiety scores were normalized to a 0-1 range. Finally, to address missingness without introducing external bias or distorting feature distributions, we applied a random imputation strategy. For any variable with missing entries, values were sampled directly from the observed non-missing distribution within the same population subgroup.

## 2.3 Predictive modeling framework

To robustly identify correlates of depression and evaluate the stability of risk factors across populations, we evaluated two complementary predictive models: logistic regression and random forest (Breiman, 2001). The use of logistic regression provided interpretable odds ratios and statistical significance testing for linear associations, while the random forest approach was employed to capture potential non-linear relationships and high-order interactions between predictors.

The modeling process was strictly stratified by population type to enable parallel evaluation. For the student cohort, the India student dataset (n = 502) was randomly partitioned into a training set (n = 401) and a test set (n = 101). The size of the internal test set was deliberately chosen to match the sample size of the external Malaysia validation cohort (n = 101), allowing for a balanced comparison of internal versus external model performance. Similarly, for the worker cohort, the large-scale Chinese dataset (n = 25,586) was split into a training set (n = 23,030) and a test set (n = 2,556). Here too, the test set size was selected to match the India workers dataset (n=2,556). This parallel setup enabled a consistent and controlled evaluation framework, where the performance of each model trained on a large base dataset could be compared against both internal (same-population) and external (cross-population) samples of equivalent size.

Logistic regression models were initially fitted using all available predictors. To optimize model parsimony and explanatory power, we subsequently performed stepwise model selection using the Akaike Information Criterion (AIC), employing both forward and backward selection procedures to retain only the most statistically significant predictors. Random forest model was constructed with 500 decision trees using default parameters for maximum depth and minimum samples per leaf. Feature importance was assessed using the Mean Decrease in Gini Impurity, quantifying the contribution of each predictor to the model's classification accuracy. Model performance for both algorithms was evaluated using prediction accuracy and confusion matrices across the training, internal testing, and external validation datasets.

## 2.4 Causal machine learning and mechanism identification

### 2.4.1 Causal Forest

To move beyond average associations and examine whether the effect of pressure on depression varies across individuals, we implemented a causal forest model (Athey et al., 2019). This nonparametric ensemble method estimates the conditional average treatment effect (CATE), defined as the expected difference in outcomes between treated and control units conditional on a set of covariates $X$:

$$\tau(x) = \mathbb{E}[Y(1) - Y(0)|X = x] \tag{1}$$

where $Y(1)$ and $Y(0)$ are the potential outcomes under high-pressure and low-pressure conditions, respectively. Unlike standard regression, which typically assumes a constant effect size across the population, the causal forest constructs honest trees by randomly splitting the data into two subsamples: one for determining the optimal splits and another for estimating treatment effects within the resulting leaves.

We treated pressure as the binary treatment variable and included all other covariates as potential moderators. The resulting CATE estimates allowed us to identify specific subgroups for whom the impact of pressure might be amplified. We applied a calibration test to the final model to verify that the estimated heterogeneity was statistically robust. A non-significant calibration test would indicate that the estimated heterogeneity patterns are consistent with the data.

### 2.4.2 Causal Mediation Analysis (CMA)

Following the heterogeneity analysis, we used Causal Mediation Analysis (CMA) to decompose the total effect of risk factors into direct and indirect pathways. This analysis was conducted within a counterfactual framework. Let $Y(t, m)$ denote the potential outcome for an individual under treatment $t$ and mediator value $m$, and let $M(t)$ be the potential mediator value under treatment $t$.

We estimated two key quantities. Average causal mediation effect (ACME) represents the portion of the effect transmitted through the mediator (indirect effect).

$$ACME(t) = \mathbb{E}[Y\big(t, M(1)\big) - Y(t, M(0))] \tag{2}$$

Average direct effect (ADE) represents the effect of the treatment on the outcome holding the mediator constant (direct effect).

$$ADE(t) = \mathbb{E}[Y(t, M(0)) - Y(t, M(0))] \quad (3)$$

We constructed mediator models using linear regression and outcome models using logistic regression. Identification of these effects relies on the sequential ignorability assumption, which posits that there are no unmeasured confounders for the treatment-outcome, treatment-mediator, or mediator-outcome relationships. We specifically tested multiple pathway configurations, by treating age, academic performance, and satisfaction as exposures, to isolate which factors operated through the pressure mechanism.

#### 2.4.3 Determination of causal directionality

A critical interpretational challenge in mediation analysis is distinguishing the true causal direction between highly correlated constructs. In our analysis, it was theoretically plausible that pressure mediates the effect of life satisfaction on depression (Satisfaction → Pressure → Depression) or, conversely, that life satisfaction mediates the effect of pressure (Pressure → Satisfaction → Depression). When we initially modeled both directions, standard CMA yielded significant indirect effects for both pathways, creating an ambiguity that could not be resolved by the mediation estimates alone.

To resolve this, we employed a rigorous two-step validation process. First, we applied residual-based independence tests (as proposed by Wiedermann and von Eye (2016)). This method is grounded in the Darmois-Skitovich Theorem (Darmois, 1953; Skitovich, 1953) and a decision rule framework that relies on testing the statistical independence between model residuals and predictors. The underlying assumption is that in the correctly specified model, residuals should be independent of the predictors, whereas in a misspecified model, particularly when the true predictor is non-normally distributed, this independence will be violated. However, these tests proved inconclusive in our dataset.

Second, to overcome this limitation, we designed a series of numerical simulation experiments representing various underlying causal structures. We generated synthetic data for seven distinct scenarios, including cases of no mediation, partial mediation, and confounding, with known ground truths (Table 2). By comparing the resulting ACME estimates from our real data against the theoretical expectations from these synthetic benchmarks, we established an empirical basis for selecting the dominant pathway.

Table 2 Summary of data generation processes for numerical simulation experiments. Variables $x$, $y$, $z$ represent the treatment, mediator, and outcome, respectively. Error terms were sampled from non-normal distributions to test model robustness.

| **Group** | **Scenario** | **Predictor & mediator generation (*x*, *y*)** | **Outcome generation (*z*)** |
|---|---|---|---|
| 1 | Independent | $x, y \sim U$ (independent) | $z = x + \epsilon$ |
| 2.1 | Confounding (sum) | $x, y \sim U$ (independent) | $z = x + y + \epsilon$ |
| 2.2 | Confounding (weighted) | $x, y \sim U$ (independent) | $z = 0.8x + 0.2y + \epsilon$ |
| 3.1 | Mediation (Unit) | $x \sim U; y = x + \epsilon$ | $z = x + y + \epsilon$ |
| 3.2 | Mediation (strong) | $x \sim U; y = 1000x + \epsilon$ | $z = x + y + \epsilon$ |
| 3.3 | Mediation (weak) | $x \sim U; y = 0.1x + \epsilon$ | $z = x + y + \epsilon$ |
| 3.4 | Mediation (micro) | $x \sim U; y = 0.001x + \epsilon$ | $z = x + y + \epsilon$ |

Note: $U$ denotes uniform sampling from the range [1, 10,000]. $\epsilon$ denotes error terms sampled with replacement from [1, 10] to introduce non-normality. $N = 3{,}000$ for all groups.

# 3 Results

## 3.1 Predictive performance and risk factors

In the student cohort, both modeling approaches consistently identified perceived pressure as the dominant predictor of depressive symptoms. The logistic regression model achieved an internal prediction accuracy of 0.76, identifying age, pressure, workload, financial stress, mental health history, and life satisfaction as statistically significant correlates (Table 3). Specifically, the odds of depression were higher among students who were younger, reported lower life satisfaction, and experienced higher levels of pressure and financial stress. The random forest model demonstrated slightly superior internal performance (accuracy = 0.81). Feature importance analysis corroborated the regression findings, ranking pressure as the most influential feature, followed by age, life satisfaction, and financial stress (Figure 1).

Table 3 Multivariable logistic regression estimates of risk factors associated with depression in the student cohort (India dataset)

| | Estimate | Std. Error | z value | p-value | Significant? |
|---|---|---|---|---|---|
| Intercept | -0.907 | 0.975 | -0.930 | 0.352 | no |
| Gender | 0.405 | 0.283 | 1.430 | 0.153 | no |
| Age | -0.157 | 0.031 | -5.029 | 4.94e-07 | yes |
| Pressure | 1.103 | 0.126 | 8.723 | <2e-16 | yes |
| Workload | 0.403 | 0.084 | 4.801 | 1.58e-06 | yes |
| Financial stress | 0.727 | 0.114 | 6.382 | 1.75e-10 | yes |
| Mental health history | 0.647 | 0.281 | 2.298 | 0.022 | yes |
| Satisfaction | -0.680 | 0.112 | -6.059 | 1.37e-09 | yes |

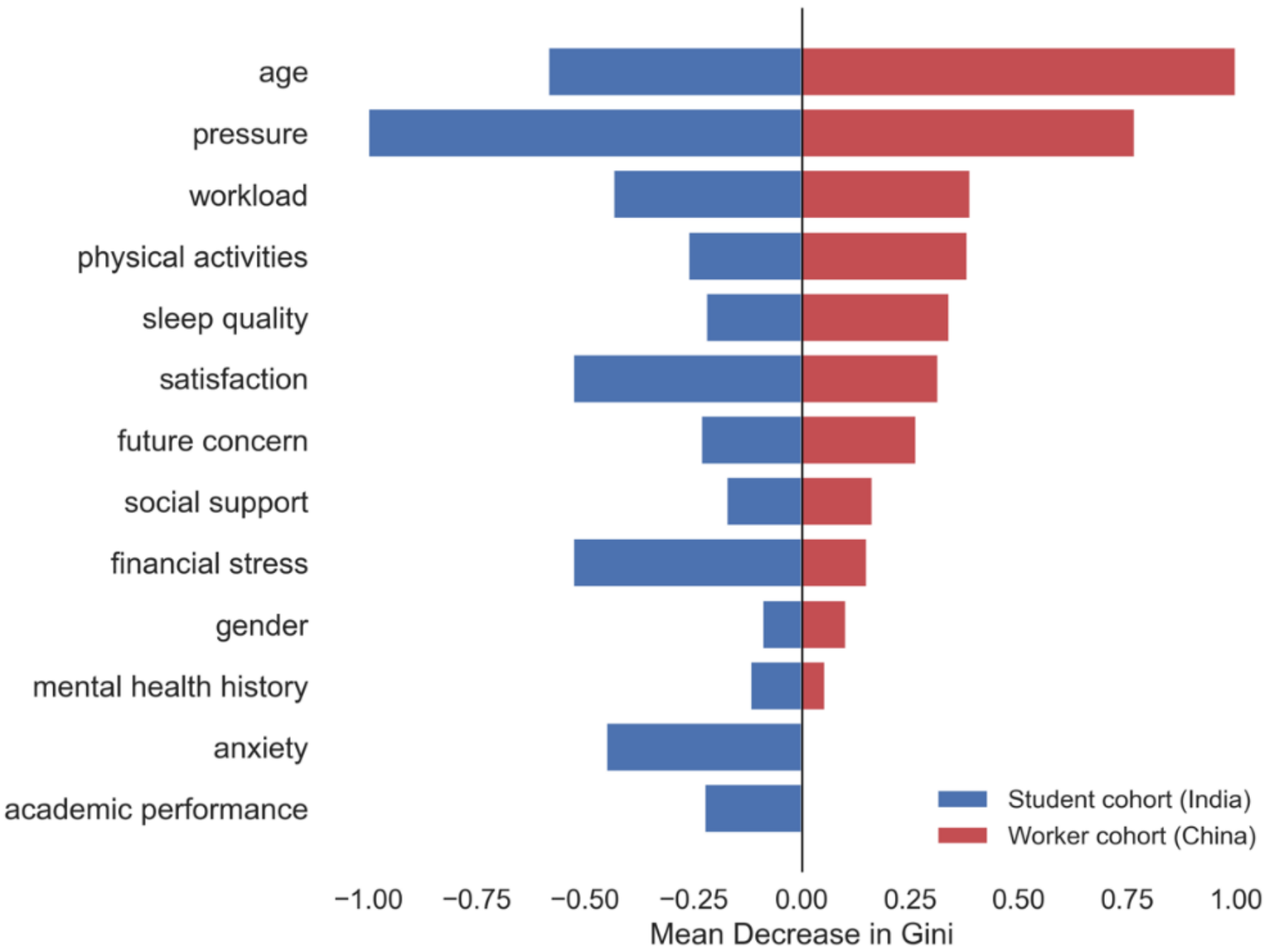

Figure 1 Comparative feature importance ranking between Student and Worker cohorts. Blue bars (left) represent the student cohort (India), and red bars (right) represent the worker cohort (China).

In the worker cohort, the risk profile exhibited significant overlap with students regarding stress factors but divergent demographic predictors. The logistic regression model achieved an internal prediction accuracy of 0.73, with significant predictors including gender, age, pressure, financial stress, mental health history, sleep quality, and satisfaction (Table 4). In contrast to the student population, depression risk in workers was positively associated with older age. The random forest model performed comparably (accuracy = 0.73) but demonstrated improved cross-national generalizability (external accuracy = 0.70 vs. 0.67 for the logistic model). Notably, the feature importance ranking for workers revealed a distinction between linear and non-linear drivers. While logistic regression identified pressure as the strongest statistical associate (highest z-value), the random forest model identified age as the leading predictor based on Gini impurity, followed by pressure and workload (Figure 1).

Table 4 Multivariable logistic regression estimates of risk factors associated with depression in the worker cohort (China dataset)

| | Estimate | Std. Error | z value | p-value | Significant? |
|---|---|---|---|---|---|
| Intercept | -0.772 | 0.156 | -4.950 | 7.40e-07 | yes |
| Gender | -0.220 | 0.032 | -6.855 | 7.14e-12 | yes |
| Age | 0.008 | 0.002 | 5.269 | 1.37e-07 | yes |
| Pressure | 0.451 | 0.010 | 44.591 | <2e-16 | yes |
| Workload | -0.014 | 0.013 | -1.105 | 0.269 | no |
| Financial stress | 0.085 | 0.032 | 2.647 | 0.008 | yes |
| Mental health history | 0.292 | 0.082 | 3.538 | 0.0004 | yes |
| Sleep quality | -0.246 | 0.013 | -18.872 | <2e-16 | yes |
| Satisfaction | -0.257 | 0.020 | -12.725 | <2e-16 | yes |
| Social support | -0.036 | 0.034 | -1.068 | 0.285 | no |
| Future concern | 0.018 | 0.013 | 1.404 | 0.160 | no |
| Physical activities | -0.003 | 0.013 | -0.216 | 0.829 | no |

A direct comparison of these cohorts reveals a distinct age-related divergence in depression risk. While pressure consistently drives depression across all groups, the influence of age follows opposite trajectories: younger age predicts higher risk in students, whereas older age predicts higher risk in the working population. This contrast highlights the context-dependent nature of demographic risk factors, suggesting that vulnerability peaks at different life stages depending on the prevailing occupational or academic environment.

Furthermore, while core predictors remained consistent, cross-national performance dropped notably during external validation. Student accuracy dropped from 0.76 to 0.55 for logistic model and 0.81 to 0.58 for random forest model; Worker accuracy from 0.73 to 0.67 for logistic model and 0.73 to 0.70 for random forest model. This decline highlights the influence of cultural and contextual factors on how depression manifests and is reported across borders. Despite these variations, perceived pressure consistently emerged as the most direct and powerful predictor across both cohorts and modeling techniques. This relationship persisted even after accounting for other variables such as satisfaction and anxiety, suggesting that stress-related mechanisms play a central, robust role in shaping mental health outcomes.

## 3.2 Causal heterogeneity

Having established pressure as a robust predictor, we applied a causal forest model to the student cohort to determine if its impact varied across individuals. The analysis produced a mean forest prediction coefficient of 1.003, indicating that the model accurately recovered the average treatment effect of pressure on depression. The subsequent calibration test suggested an absence of broad treatment-effect heterogeneity, implying that the detrimental effect of pressure was largely stable across the general student population rather than being limited to specific clusters.

However, despite this overall homogeneity, examination of the CATE revealed meaningful variations in the magnitude of vulnerability. Several covariates emerged as significant moderators of the pressure-depression relationship. Specifically, student with higher levels of anxiety (CATE = 0.221) and lower levels of life satisfaction (CATE = 0.168) exhibited the highest sensitivity to pressure (Table 5). Academic performance (CATE = 0.109) and age (CATE = 0.099) also showed elevated conditional effects. These findings indicate that while pressure is a universal risk factor, its depressogenic impact is amplified in individuals who are already psychologically vulnerable (high anxiety, low satisfaction) or academically struggling.

Table 5 Estimated Conditional Average Treatment Effects (CATE) from Causal Forest with pressure as treatment

| Variable | CATE estimate |
|---|---|
| Anxiety | 0.221 |
| Satisfaction | 0.168 |
| Academic performance | 0.109 |
| Age | 0.099 |
| Workload | 0.079 |
| Future concern | 0.065 |
| Social support | 0.059 |
| Financial stress | 0.058 |
| Extracurricular activities | 0.050 |
| Sleep quality | 0.045 |
| Mental health_history1 | 0.026 |
| Gender | 0.011 |

## 3.3 Mediation pathways and directionality

To determine whether the identified risk factors influence depression directly or operate through the mechanism of perceived pressure, we conducted the CMA. We specifically examined age, academic performance, anxiety, and life satisfaction, which were identified as high-priority moderators in the previous heterogeneity analysis. The decomposition of their effects is summarized in Table 6.

For academic performance and anxiety, the analysis indicated no significant causal transmission to depression in this framework. Despite being identified as vulnerability factors in the CATE analysis, neither variable showed a significant total or indirect effect ($p > 0.05$). This suggests that while poor grades or high anxiety may amplify the damage caused by pressure (moderation), they do not function as independent drivers of depression (mediation) in the absence of other stressors.

For age, relationship with depression was found to be almost entirely direct. The direct effect was statistically significant (ADE = -0.009, $p < 0.001$), while the mediation effect through pressure was negligible and non-significant ($p = 0.648$). This confirms that the protective effect of older age in the student cohort operates independently of perceived pressure.

Table 6 Decomposition of total, direct, and indirect effects on depression. Estimates represent the probability change associated with the predictor.

| Predictor | Total Effect | Indirect Effect via Pressure (ACME) | Average Direct Effect (ADE) | Mediation Status |
|---|---|---|---|---|
| Academic Performance | -0.005 (p=0.750) | -0.004 (p=0.662) | -0.001 (p=0.938) | No significant effect |
| Age | -0.009 (p<2e-16) | 0.0004 (p=0.648) | -0.009 (p<2e-16) | Direct only |
| Anxiety | 0.077 (p=0.212) | 0.015 (p=0.584) | 0.061 (p=0.252) | No significant effect |
| Life satisfaction | -0.089 (p<2e-16) | -0.013 (p=0.026) | -0.075 (p<2e-16) | Partial mediation |

In contrast, life satisfaction exhibited a complex, dual pathway. Both the direct effect (ADE = -0.075, $p < 0.001$) and the indirect effect through pressure (ACME = -0.013, $p = 0.026$) were statistically significant. The estimated proportion mediated was 15.1%, indicating that while life satisfaction functions primarily as a direct protective factor, a meaningful subset of its influence operates by buffering perceived pressure.

However, the bi-directional nature of the relationship between pressure and satisfaction posed an interpretational challenge. When modeled in reverse (testing if Satisfaction mediates Pressure), the indirect effect also appeared significant ($p = 0.014$), creating statistical ambiguity. To resolve this, we validated our directionality test using the synthetic datasets described in Section 2.4.3.

The results, summarized in Table 7, illustrate both the capabilities and the challenges of the CMA approach. In the unit mediation scenario (Group 3.1), both the forward and reverse models yielded estimates close to 1.0, confirming that when causal weights are balanced, standard mediation estimates can be ambiguous. However, the simulation revealed that misspecified models produce drastic deviations when the mediator is not unit weighted. For example, in Group 3.4 (Micro mediator), where the true mediator effect was small ($y = 0.001x$), the correct model ($x \rightarrow y \rightarrow$ z) accurately recovered the ACME ($\approx 0.001$), whereas the reverse model ($y \rightarrow x \rightarrow$ z) produced an implausibly large estimate ($\approx 505.0$). This discrepancy provides the empirical signature needed to identify the correct causal flow: while valid models consistently align with data structures, invalid models yield inflated or unstable estimates in non-unit scenarios.

Table 7 Validation of CMA estimates across simulated causal structures.

| Scenario (ground truth) | Expected ACME for x-y-z | Observed ACME for x-y-z (forward) | Observed ACME for y-x-z (backward) |
|---|---|---|---|
| 1 Independent | ≈ 0 (no mediation) | -3.47E-08 | 0.011 |
| 2.1 Confounding (sum) | ≈ 0 (no mediation) | 0.011 | 0.011 |
| 2.2 Confounding (weighted) | ≈ 0 (no mediation) | 0.002 | 0.009 |
| 3.1 Mediation (unit) | ≈ 1 | 1.007 | 0.993 |
| 3.2 Mediation (strong) | ≈ 1000 | 1006.741 | -0.006 |
| 3.3 Mediation (weak) | ≈ 0.1 | 0.101 | 9.997 |
| 3.4 Mediation (micro) | ≈ 0.001 | 0.001 | 505.002 |

We then applied this benchmark to the India student dataset to resolve the ambiguity between pressure and satisfaction. We compared the observed ACME values against the theoretical expectations derived from the simulation benchmarks. the observed indirect effect for the Satisfaction→Pressure→Depression pathway (Observed ACME = -0.128) aligned closely with the theoretical expectation derived from the data structure (Expected ≈ -0.109) (Table 8). Conversely, the reverse pathway showed a substantial deviation between observed and expected values. This empirical validation provides robust evidence that the dominant causal flow runs from Life satisfaction to Pressure, confirming that pressure acts as the proximal mediator through which deficits in life satisfaction partially translate into depressive symptoms.

Table 8 Directionality determination for the India student cohort.

| | Expected ACME | Observed ACME | Conclusion |
|---|---|---|---|
| Pressure→Satisfaction→Depression | ≈ -0.105 | 0.075 | Rejected |
| Satisfaction→Pressure→Depression | ≈-0.109 | -0.128 | Accepted |

# 4 Discussion

Our finding that pressure directly influences depression, while life satisfaction affects depression primarily through its impact on pressure, advances current understanding of depression risk mechanisms by clarifying how these factors operate together rather than in isolation. Prior research has often treated pressure and satisfaction as parallel correlates of depression, leaving open questions about their causal interplay and relative positioning within broader risk pathways. By applying CMA alongside predictive modeling approaches (logistic regression and random forests), we provide converging evidence that validated by simulation experiments and that pressure functions as a strong, proximal driver of depression, whereas life satisfaction contributes indirectly by shaping individuals' perceived stress levels (proportion mediated ≈ 15.1%). This mechanistic insight helps explain why interventions focused solely on enhancing life satisfaction may yield limited mental health benefits unless they also address underlying stressors that precipitate depressive symptoms.

Consistent with this interpretation, existing intervention research underscores the importance of directly targeting stress processes. For example, mindfulness-based stress reduction (MBSR) programs have been shown in randomized trials to reduce depression and anxiety symptoms by improving emotional regulation and attenuating the effects of stress on psychological functioning, indicating that modifying stress responses can directly alleviate depressive symptoms (Wu et al., 2025). Beyond mindfulness-based approaches, stress management training and coping skills programs have been shown to enhance psychological resilience and self-efficacy, which in turn buffer the impact of stress on mental health outcomes (Bahraseman et al., 2021). Additionally, programs that encourage adaptive coping mechanisms and enhance life satisfaction, such as cognitive-behavioral skill building, contribute to lower depressive symptoms by improving individuals' responses to stress. These findings align with conceptual models in which life satisfaction influences depression through stress appraisal and coping processes rather than acting as a direct protective factor (Almeida et al., 2021).

Building on these mechanistic insights, the present study also identified several significant predictors of depression that were consistent across both student and worker populations, including pressure, workload, financial stress, mental health history, age, and life satisfaction. These results largely replicate and reinforce prior findings showing that stress exposure, financial strain, and subjective well-being are central determinants of depressive symptoms across Asian contexts (Lowe et al., 2009; Singh et al., 2017; Islam et al., 2022). Importantly, by applying logistic regression and random forest models in parallel, our analyses demonstrate that these predictors remain robust across different modeling frameworks. The consistent prominence of pressure across models and populations further emphasizes its central role in depression risk, increasing confidence that the observed associations reflect core dimensions of vulnerability rather than artifacts of a particular analytic approach.

Beyond confirming established risk factors, a key contribution of this study lies in its direct comparison of students and workers using harmonized variables and parallel analytical frameworks. This design enabled us to identify a meaningful divergence in the role of age across populations: younger individuals were more vulnerable to depression within the student cohort, whereas older individuals faced higher risk in the worker cohort. Prior studies have often examined age effects within a single population, such as students, working adults, or older adults, leading to mixed or context-dependent conclusions about whether depression risk increases or decreases with age. For example, analyses of the National Health and Nutrition Examination Survey (NHANES) report that depression prevalence is highest among adolescents and declines with increasing age in the general U.S. population, with particularly elevated rates among individuals aged 12-19 (Brody et al., 2025). Our findings suggest that age does not exert a uniform influence on depression risk; rather, its effect is contingent on life stage and contextual stress exposure. Also, our findings caution against generalizing demographic risk factors across populations: a variable that is protective in one context (aging in university) becomes a risk factor in another (aging in the workforce).

Interpreted through this lens, the observed age-related divergence appears consistent with population-specific stressors. Among students, younger age may be associated with heightened academic pressure, transitional challenges, and limited coping resources, increasing vulnerability to depressive symptoms. In contrast, among workers, older age may reflect cumulative occupational stress, job insecurity, declining physical health, or increasing financial and family responsibilities. Because most existing studies focus on a single population at a time, such cross-population contrasts have been difficult to observe. By placing students and workers within a unified comparative framework, this study clarifies how age interacts with contextual stressors to shape depression risk and underscores the importance of population-specific interpretations of commonly used predictors.

Furthermore, our causal forest analysis refined this risk profile by identifying sources of heterogeneity within the student population. While pressure was a universal risk factor, we observed that its impact was significantly amplified in students with high anxiety (CATE=0.221) and low life satisfaction (CATE=0.168). This interaction suggests a synergistic toxicity where pre-existing psychological vulnerabilities sensitize individuals to the effects of environmental pressure. Clinically, this advocates for precision screening protocols that do not merely assess stress load but also identify students with high trait anxiety, as they represent the subgroup for whom stress-reduction interventions will yield the highest preventative benefit.

Taken together, these findings highlight the value of integrating causal inference with predictive analytics in mental health research. While predictive models are effective for identifying correlates of depression, they provide limited guidance for intervention design without insight into how risk factors exert their effects. By incorporating CMA, we move beyond identifying predictors to illuminating mechanistic pathways. Specifically, how life satisfaction and pressure jointly shape depressive outcomes. This integrated approach offers both theoretical and practical contributions, providing a stronger foundation for prevention and treatment strategies that directly address stress processes and their life-course contexts.

At the same time, several limitations warrant consideration. First, our analysis relies on cross-sectional data. While our numerical experiments and sensitivity analyses robustly support the proposed causal structure, definitive temporal ordering requires longitudinal follow-up. Second, while we performed detailed causal heterogeneity and mediation analyses for the student cohort, parallel analyses were not conducted for the worker cohort due to data harmonization constraints. Key psychological moderators identified in the student analysis, specifically anxiety and academic performance, were not available in the harmonized worker dataset (CHARLS). Consequently, we restricted the causal forest applications to the student population where these granular psychometric variables were fully observed. Future research should build on this work by employing longitudinal or experimental designs to more rigorously evaluate temporal ordering and to verify whether reductions in pressure indeed mediate the beneficial effects of improved life satisfaction on depression over time.

## 5 Conclusions

This study demonstrates the power of integrating causal machine learning with predictive analytics to move beyond identifying who is at risk of depression to understanding why risks emerge. We examined predictors of depression across student and worker populations in India, Malaysia, and China, highlighting both shared influences and context-specific patterns. Across all groups, pressure emerged as the central and most consistent predictor. Our simulation-validated mediation analysis resolves a critical theoretical ambiguity, confirming that life satisfaction mitigates depression in part by lowering perceived pressure, rather than functioning exclusively as a direct protective mechanism. Furthermore, our cross-population comparison reveals an interesting developmental reversal: age is not a uniform risk factor but interacts dynamically with life stage, predicting vulnerability in younger students yet resilience in older workers. Specifically, clinical and institutional support systems should prioritize direct stress reduction and target individuals with high trait anxiety, as addressing these proximal drivers offers the most effective pathway to breaking the causal cycle of depression.

## Data Availability

All datasets supported by this study are publicly available from Kaggle and CHARLS. The India cohort data (Depression Survey/Dataset for Analysis) is available at https://www.kaggle.com/datasets/sumansharmadataworld/depression-surveydataset-for-analysis (accessed on January 11, 2026); the Malaysia cohort data (Student Mental Health) is available at https://www.kaggle.com/datasets/shariful07/student-mental-health/data (accessed on January 11, 2026); and the China cohort data is available at https://charls.charlsdata.com/pages/Data/harmonized_charls/en.html (accessed on January 11, 2026).

## CRediT authorship contribution statement

Zhaojin Nan: Formal analysis, Investigation, Methodology, Validation, Visualization, Writing - original draft, Writing - review and editing; Ran Chen: Conceptualization, Methodology, Resources, Supervision, Writing - review and editing.

## Funding

This research was supported by a cash funding from the Office of Undergraduate Research at Washington University in St. Louis.

## Declaration of competing interest

No potential conflicts of interest relevant to this article were reported.

## Acknowledgements

I gratefully acknowledge Washington University in St. Louis and the Office of Undergraduate Research for financially supporting this project.

## References

Abécassis, J., Dumas, É., Alberge, J., Varoquaux, G., 2025. From prediction to prescription: machine learning and causal inference for the heterogeneous treatment effect. *Annual Review of Biomedical Data Science*. https://doi.org/10.1146/annurev-biodatasci-103123-095750

Almeida, D., Monteiro, D., Rodrigues, F., 2021. Satisfaction with life: mediating role in the relationship between depressive symptoms and coping mechanisms. *Healthcare* 9, Article 787. https://doi.org/10.3390/healthcare9070787

Arslan, G., Jarden, A., Yıldırım, M., Jarden, R., Silapurem, L., 2025. Longitudinal associations between life events, hope, life satisfaction, happiness, and depressive symptoms. *Cognitive Therapy and Research*. https://doi.org/10.1007/s10608-025-10630-0

Athey, S., Tibshirani, J., Wager, S., 2019. Generalized random forests. *Annals of Statistics*. https://doi.org/10.1214/18-AOS1709

Bahraseman, Z.G., Shahrbabaki, P.M., Nouhi, E., 2021. The impact of stress management training on stress-related coping strategies and self-efficacy in hemodialysis patients: a randomized controlled clinical trial. *BMC Psychiatry* 9, Article 177. https://doi.org/10.1186/s40359-021-00678-4

Besharat, M.A., Issazadegan, A., Etemadinia, M., Golssanamlou, S., Abdolmanafi, A., 2014. Risk factors associated with depressive symptoms among undergraduate students. *Asian Journal of Psychiatry* 10, 21-26. https://doi.org/10.1016/j.ajp.2014.02.002

Bonde, J.P., 2008. Psychosocial factors at work and risk of depression: a systematic review of the epidemiological evidence. *Occupational and Environmental Medicine* 65 (7), 438-445. https://doi.org/10.1136/oem.2007.038430

Breiman, L., 2001. Random forests. *Machine Learning* 45 (1), 5-32. https://doi.org/10.1023/A:1010933404324

Brody, D.J., Hughes, J.P., 2025. Depression prevalence in adolescents and adults: United States, August 2021-August 2023. *NCHS Data Brief* (527), 1-8. https://doi.org/10.15620/cdc/174579

Darmois, G., 1953. Analyse générale des liaisons stochastiques: étude particulière de l'analyse factorielle linéaire. *Revue de l'Institut International de Statistique* 21 (1-2), 2-8. https://doi.org/10.2307/1401511

Duan, J., Wang, X., Yu, W., Deng, Y., Tu, Q., Giri, M., Xiao, M., Lv, Y., 2020. Risk factors associated with depressive symptoms in older people based on comprehensive geriatric assessment in Chongqing, China. *Research Square*. https://doi.org/10.21203/rs.3.rs-40283/v1

Islam, S., Akter, R., Sikder, T., Griffiths, M.D., 2022. Prevalence and factors associated with depression and anxiety among first-year university students in Bangladesh: a cross-sectional study. *International Journal of Mental Health and Addiction* 20, 1289-1302. https://doi.org/10.1007/s11469-020-00242-y

Islam, S., 2023. Student Mental Health [dataset]. *Kaggle*. https://www.kaggle.com/datasets/shariful07/student-mental-health

Imai, K., Keele, L., Tingley, D., 2010. A general approach to causal mediation analysis. *Psychological Methods*. https://doi.org/10.1037/a0020761

Lee, J., Kim, E., Wachholtz, A., 2016. The effect of perceived stress on life satisfaction: the mediating effect of self-efficacy. *Chongsonyonhak Yongu*. https://doi.org/10.21509/KJYS.2016.10.23.10.29

Li, R., Wang, X., Luo, L., Yuan, Y., 2024. Identifying the most crucial factors associated with depression based on interpretable machine learning: a case study from CHARLS. *Frontiers in Psychology* 15, Article 1392240. https://doi.org/10.3389/fpsyg.2024.1392240

Lin, Y.C., Huang, S.S., Yen, C.W., Kabasawa, Y., Lee, C.H., Huang, H.L., 2022. Physical frailty and oral frailty associated with late-life depression in community-dwelling older adults. *Journal of Personalized Medicine*. https://doi.org/10.3390/jpm12030459

Liu, A., Peng, Y., Zhu, W., Zhang, Y., Ge, S., Zhou, Y., Zhang, K., Wang, Z., He, P., 2021. Analysis of factors associated with depression in community-dwelling older adults in Wuhan, China. *Frontiers in Aging Neuroscience* 13, Article 743193. https://doi.org/10.3389/fnagi.2021.743193

Lowe, G., Lipps, G., Young, R., 2009. Factors associated with depression in students at the University of the West Indies, Mona, Jamaica. *West Indian Medical Journal* 58, 21-27.

Ogden, R., Schoetensack, C., Klegr, T., Pestana, J.V., Valenzuela, R., Goncikowska, K., Giner-Domínguez, G., Papastamatelou, J., Chappuis, S., Fernández Boente, M., Meteier, Q., Černohorská, V., Codina, N., Martin-Söelch, C., Wittmann, M., Witowska, J., 2025. Chronic time pressure as a predictor of symptoms of depression, anxiety and stress. *BMC Psychology* 13, Article 1407. https://doi.org/10.1186/s40359-025-03654-4

Ooi, P.B., Khor, K.S., Tan, C.C., Ong, D.L.T., 2022. Depression, anxiety, stress, and satisfaction with life: moderating role of interpersonal needs among university students. *Frontiers in Public Health* 10, Article 958884. https://doi.org/10.3389/fpubh.2022.958884

Padmanabhanunni, A., Pretorius, T.B., Isaacs, S.A., 2023. Satisfied with life? The protective function of life satisfaction in the relationship between perceived stress and negative mental health outcomes. *International Journal of Environmental Research and Public Health* 20, Article 6777. https://doi.org/10.3390/ijerph20186777

Rissanen, T., Viinamäki, H., Honkalampi, K., Lehto, S.M., Hintikka, J., Saharinen, T., Koivumaa-Honkanen, H., 2011. Long-term life dissatisfaction and subsequent major depressive disorder and poor mental health. *BMC Psychiatry* 11, Article 140.

Sharma, S., Rao, R., Gujral, D., Pathak, S., 2025. Depression Survey/Dataset for Analysis [dataset]. *Kaggle*. https://www.kaggle.com/datasets/sumansharmadataworld/depression-surveydataset-for-analysis

Singh, M.M., Gupta, M., Grover, S., 2017. Prevalence and factors associated with depression among schoolgoing adolescents in Chandigarh, North India. *Indian Journal of Medical Research*, 205-215. https://doi.org/10.4103/ijmr.IJMR_1339_15

Skitovich, V.P., 1953. On a property of the normal distribution. *Doklady Akademii Nauk SSSR* 89, 217-219.

Wang, X., Cao, X., Yu, J., Jin, S., Li, S., Chen, L., Liu, Z., Ge, X., Lu, Y., 2024. Associations of perceived stress with loneliness and depressive symptoms: the mediating role of sleep quality. *BMC Psychiatry* 24, Article 172. https://doi.org/10.1186/s12888-024-05609-2

Wiedermann, W., von Eye, A., 2016. Directionality of effects in causal mediation analysis. In: *Statistics and Causality: Methods for Applied Empirical Research*, pp. 63-106. https://doi.org/10.1002/9781118947074.ch4

Worrall, C., Jongenelis, M.I., McEvoy, P.M., Jackson, B., Newton, R.U., Pettigrew, S., 2020. An Exploratory Study of the Relative Effects of Various Protective Factors on Depressive Symptoms Among Older People. *Frontiers in Public Health* 8, Article 579304. https://doi.org/10.3389/fpubh.2020.579304

Wu, Y., Fan, L., Xia, F., Zhou, Y., Wang, H., Feng, L., Xie, S., Xu, W., Xie, Z., He, J., Liu, D., He, S., Xu, Y., Deng, J., Wang, T., Chen, L., 2024. Global, regional, and national time trends in incidence for depressive disorders from 1990 to 2019: an age-period-cohort analysis for the GBD 2019. *Annals of General Psychiatry* 23, Article 28. https://doi.org/10.1186/s12991-024-00513-1

Wu, Y., Ban, Y., Pan, G., Yao, M., Liu, L., Chen, T., Wu, H., 2025. Effects of online mindfulness-based stress reduction training on depression and anxiety symptoms among psychiatric healthcare workers in a randomized controlled trial: the mediating role of emotional suppression. *BMC Psychiatry* 25, Article 577. https://doi.org/10.1186/s12888-025-06967-1

Zhao, Y., Hu, Y., Smith, J.P., Strauss, J., Yang, G., 2014. The China Health and Retirement Longitudinal Study (CHARLS). *International Journal of Epidemiology* 43 (1), 61-68. https://doi.org/10.1093/ije/dys203

## Figures:

Figure 1 Comparative feature importance ranking between Student and Worker cohorts. Blue bars (left) represent the student cohort (India), and red bars (right) represent the worker cohort (China).

## Tables:

Table 1 Summary of study cohorts. GPA: current grade point average; CHARLS: the Harmonized China Health and Retirement Longitudinal Study

Table 2 Summary of data generation processes for numerical simulation experiments. Variables x, y, z represent the treatment, mediator, and outcome, respectively. Error terms were sampled from non-normal distributions to test model robustness.

Table 3 Multivariable logistic regression estimates of risk factors associated with depression in the student cohort (India dataset)

Table 4 Multivariable logistic regression estimates of risk factors associated with depression in the worker cohort (China dataset)

Table 5 Estimated Conditional Average Treatment Effects (CATE) from Causal Forest with pressure as treatment

Table 6 Decomposition of total, direct, and indirect effects on depression. Estimates represent the probability change associated with the predictor.

Table 7 Validation of CMA estimates across simulated causal structures.

Table 8 Directionality determination for the India student cohort.